\documentstyle[psfig]{mn}
\def\apgt{\ {\raise-.5ex\hbox{$\buildrel>\over\sim$}}\ }
\def\aplt{\ {\raise-.5ex\hbox{$\buildrel<\over\sim$}}\ }

\def\pe{\mbox{$P_{\rm orb} - e$}~}
\def\m{^m\kern-7pt .\kern+3.5pt}
\newcommand{\orb}{\mbox{$_{\rm orb}$}}
\newcommand{\msun}{\mbox{${\rm M}_\odot$}}
\newcommand{\rsun}{\mbox{${\rm R}_\odot$}}

\def\unit#1{{\mbox{[{\rm #1}]}}}
\def\unit#1{{\mbox{[{\rm #1}]}}}

\newcommand{\ms}{\mbox{${\it ms}$}}

\newcommand{\ns}{\mbox{${\it ns}$}}
\newcommand{\wid}{\mbox{${\it wd}$}}
\newcommand{\bh}{\mbox{${\it bh}$}}

\def\etal{{et al.}\ }
\newcommand{\md}{\mbox {$\dot{M}$}}
\newcommand{\myr}{\mbox {~${\rm M_{\odot}~yr^{-1}}$}}
\newcommand{\pyr}{\mbox {{\rm yr$^{-1}$}}}
\def\eg{{\it e.g.}\ }
\newcommand{\psr}{\mbox{PSR~B1820--11}}
\newcommand{\psrb}{\mbox{PSR~B2303+46}}
\def\apj{ApJ}

\title
{The possible companions of young radio pulsars}

\author[S.F.~Portegies~Zwart and L.R.~Yungelson]
{Simon~F.~Portegies~Zwart$^{1,2}$\thanks{Hubble Fellow}
and
Lev R. Yungelson$^3$ \\
$^1$Department of Astronomy
	    Boston University,
	    725 Commonwealth Avenue,
	    Boston, MA 01581, USA; spz@komodo.bu.edu\\
$^2$ Department of Information Science and Graphics, 
	College of Arts and Science, 
	University of Tokyo, \\ 
        3-8-1 Komaba, 
	Meguro-ku, Tokyo 153, Japan \\
$^3$Institute of Astronomy of the Russian Academy of Sciences, 48 
Pyatnitskaya Str., 109017 Moscow, Russia; lry@inasan.rssi.ru\\
}

\date{Accepted 1999.
      Received 1998;
      in original form 1998}

\pagerange{\pageref{firstpage}--\pageref{lastpage}}
\pubyear{1999}

\begin{document}
\maketitle
\label{firstpage}
\begin{abstract}
We discuss the formation of pulsars with massive companions in
eccentric orbits.  We demonstrate that the probability for a
non-recycled radio pulsar to have a white dwarf as a companion is
comparable to that of having an old neutron star as a companion.
Special emphasis is given to \psr\ and PSR B2303+46.  Based on
population synthesis calculations we argue that \psr\ and PSR B2303+46
could very well be accompanied by white dwarfs with mass $\apgt
1.1$\,\msun.  For \psr, however, we can not exclude the possibility
that its companion is a main-sequence star with a mass between $\sim
0.7$\,\msun\ and $\sim 5$\,\msun.

\end{abstract}

\section{Introduction}

High mass binary pulsars are binaries in which a radio pulsar is
accompanied by an unseen companion that is approximately as massive as
the pulsar. Although the companion is generally believed to be another
neutron star, it appears to be likely that some of the high-mass
binary pulsars are not accompanied by neutron stars but are instead
accompanied by white dwarfs or low-mass main-sequence stars.

PSR B1820-11 (J1823-1115), for example, is a radio pulsar in a binary
with an orbital period $P\orb \approx 357.8$~days and an eccentricity
$e \approx 0.795$~(Lyne \& McKenna 1989).  The derived mass function
$f = 0.068\,\msun$ indicates that the mass of the companion exceeds
$\sim 0.7\,\msun$.  The small spin-down age of the pulsar ($\tau =
6.5$\,Myr) and its strong magnetic field ($ B \simeq 6 \times 10^{11} {\rm
G}$) suggest that it is a young -- non-recycled -- pulsar (Phinney \&
Verbunt 1991).

Based on the mass function and the orbital eccentricity, Lyne \&
McKenna (1989) suggest that the most likely companion to \psr\ is a
neutron star. The companion is unlikely to be a massive hydrogen star
or a black hole due to the low mass function.  Lyne \& McKenna argue
that the observed orbital eccentricity excludes the companion as a
white dwarf or a low-mass helium star.  Phinney \& Verbunt (1991)
suggest that the companion of \psr\ may be a $\sim 1$\,\msun\
main-sequence star, which will make the system a precursor of low-mass
X-ray binary.

Also PSR B2303+46 may contain a non-recycled pulsar ($P = 1.06$\,s,
$\tau = 30$\,Myr, $B=8 \times 10^{11}$\,G) in a binary with $P_{\rm
orb} = 12.34$\,day and $e \approx 0.66$ (Stokes et al. 1985).  The
total mass of the binary is $2.64 \pm 0.05$\,\msun (Thorsett \&
Chakrabarty 1999) and no optical counterpart was detected down to $R =
26\m$ (Kulkarni 1988). Recently van Kerkwijk \& Kulkarni (1999),
however, discovered an object which coincides with the timing position
of \psrb\ and the properties of this object are consistent with those
of a massive white dwarf with a cooling age close to the age of
pulsar.

With population synthesis calculations we demonstrate that the birth
rate of binaries in which the neutron star is born {\em after} the
white dwarf is comparable to the birth rate of binaries with two
neutron stars. In the former case the binary will have a high
eccentricity eccentric and the radio pulsar will be accompanied by a
white dwarf with a mass $\apgt 1.1$\,\msun. In this paper we argue
that both \psr\ and \psrb\ systems may be formed via such a scenario
(see \S\,2.2).  The calculations, however, do not rule out that the
companion of \psr\ is either another neutron star or a main-sequence
star.  In the latter case, the mass of the main-sequence star would
most likely be between $\sim 0.7\,\msun$ and $\sim 5$\msun.

\section{The binary companions of \psr\ and PSR B2303+46}

\subsection{Hiding a main-sequence star}
\label{secms}

At a distance of 6.3\,kpc and a height above the Galactic plane of
$\sim 100$\,pc (Taylor, Manchester, \& Lyne 1993), \psr\ is heavily
obscured by the interstellar medium. A star with an absolute magnitude
of $M_{\rm V} \sim -4\m0$ could easily be missed in a survey with a
limiting magnitude of $20\m0$ (assuming an interstellar absorption of
$A_{\rm V} = 1\m6\,{\rm kpc^{-1}})$.  In the USNO-A v1.0 catalog
(Monet \etal\ 1997), based on the Palomar Sky Survey, the object
closest to \psr\ is located at a distance of 5.9 seconds of arc and
therefore cannot be associated with \psr.  There is no indication for
other forms of radiation (X-rays, infrared, etc.) from the direction
of \psr.  These suggest that the companion can not be a very massive
star or have a strong stellar wind.

The presence of radio emission indicates that the rate of mass loss by
the companion star is modest.  An upper limit to \md\ in the stellar
wind of the companion can be estimated from the expression for
the optical depth of the stellar wind for free-free absorption (\eg,
Eq.~16 in Illarionov \& Sunyaev 1975).  By assuming that the wind of
the accompanying star is transparent at $\lambda = 75$\,cm (and a
temperature of the stellar plasma of $10^4$\,K) we obtain
\begin{equation}
\md \aplt 4.8\,10^{-12} {M_{\rm tot} \over [\msun]} \, {P\orb \over
	[{\rm day}]} \,
	(1-e)^{\frac{3}{2}} \,\,\; [\myr],
\label{eq:ppulsar}
\end{equation}
where $M_{\rm tot}$ is the mass of the binary.  The estimate for \md\
depends on the eccentricity; substitution of $(1+e)$ for $(1-e)$ in
Eq.~(\ref{eq:ppulsar}) gives the lower limit for the mass loss rate
given that the radio pulsar is only visible at apocenter. We
illustrate this limit in Fig.\, 1 for two typical combinations of $M_{\rm
tot}$\ and \md.

A binary with more or less similar orbital characteristics is
PSR~B1259-63 ($P\orb = 1237\pm24$\,days, $e=0.8699,\ M_{opt}= 10 \pm 3
\msun, \md = 5 \times 10^{-8} \myr$; Johnston \etal 1992, 1994) which
contains a radio pulsar and a Be-type star.  The pulsar is visible at
apastron but not at periastron. This is understood from
Eq.\,(\ref{eq:ppulsar}) as an effect of shielding of the radio signal
from the pulsar by the stellar wind.  Another known radio pulsar in a
long-period binary is PSR J0045-7319 ($P\orb = 147.8 $\,days, $e=
0.808$) with a B1 V ($M = 8.8 \pm 1.8$\,\msun) companion with $\md <
3.4 \times 10^{-11} (v_{\inf}/v_{\rm esc})$\,\myr\ (McConnell \etal
1991; Bell \etal 1995; Kaspi, Tauris \& Manchester 1996). This pulsar
is visible along its entire orbit due to a more tenuous wind of the
companion star, which is typical for a lower metallicity star in the
Large Magellanic Cloud.

The absence of eclipses and of X-rays from accretion or shocks
generated by the interaction between the pulsar wind with wind of the Be star
(as are observed in PSR B1259-63, Grove \etal\ 1995) also indicates that the
companion of \psr\ is not likely to be a massive star.  An upper limit
to the mass-loss rate of the companion star in \psr\ may be set by
measuring variations in the dispersion measure of the radio pulsar
when it is close to periastron (see Melatos, Johnston, \& Melrose
1995).

\subsection{Can the companion to a radio pulsar be a massive white dwarf?}
\label{secwd} 

The eccentricity of a binary orbit can either be primordial or induced
by the sudden mass loss by one of the stars (Blaauw 1961). The latter
is expected to occur in a supernova in which the exploding star loses
a considerable fraction of its mass leaving behind a neutron star.
The orbital periods of \psr\ and \psrb\ prohibit the observed
eccentricity from being primordial; these binaries have experienced a
phase of mass transfer which circularized the orbit.  The orbital
eccentricities of these binaries are therefore induced by a supernova
which may have produced the currently observed pulsar.

If the companions to \psr\ or PSR B2303+46 are white dwarfs they must
have been formed {\em before} the neutron stars.  This possibility was
never studied in detail since the progenitor of a white dwarf is
expected to live longer than the progenitor of a neutron star.
However, it was noticed by Tutukov \& Yungelson (1993) and by
Portegies Zwart \& Verbunt (1996) that a reversal of evolution can be
accomplished if both stars are similar in mass and if the masses of
both stars are only slightly smaller than the mass limit for forming a
neutron star.  If this is the case, the secondary star may still gain
enough mass in a phase of mass transfer to pass the limit for forming
a neutron star.  For this to happen mass transfer should proceed
rather conservatively, which will be the case since the initial mass
ratio is close to unity.  In the second, unstable, phase of mass
transfer the white dwarf (original primary star) will spiral-in into
the envelope of the secondary as this star ascends the giant branch.
This phase causes the orbital period to decrease dramatically by using
orbital energy to carry the common envelope to infinity.  If the white
dwarf and the core of the giant stay detached after the common
envelope is ejected, a close binary consisting of a white dwarf and a
helium star remains.

If the helium star is massive enough, the binary experiences a
supernova and a young radio pulsar is formed. If the system survives
the supernova the companion of the radio pulsar will be a white dwarf
in an eccentric orbit.  The mass of the white dwarf will be $\apgt
1.1$\msun in such a case, because it originated from a star that was
rather massive.

\subsection{Binary population synthesis}

We use the population synthesis program for binary stars {\sf
SeBa}\footnote{The name {\sf SeBa} is adopted from the Egyptian word
for `to teach', `the door to knowledge' or `(multiple) star'. The
exact meaning depends on the hieroglyphic spelling.}  (Portegies Zwart
\& Verbunt 1996) for evolving a million binaries with a primary mass
between 8\,\msun\ and 100\,\msun. We used model B from Portegies Zwart
\& Yungelson (1998; henceforth PZY98), which satisfactorily reproduces
the properties of observed high-mass binary pulsars (with neutron star
companions). For a single star we adopt a minimum zero-age mass of
8\,\msun\ for forming a neutron stars, which coincides with estimates
based on observations (Koester \& Reimers 1996, see however Ritossa,
Garcia-Berro, \& Iben (1996) who show that a single star with a mass
of 10\,\msun\ may still evolve into a oxygen-neon-magnesium white
dwarf).  If a star is stripped of its envelope in an early phase this
lower limit may become as large as 12\,\msun\footnote{In {\sf SeBa}
the lower limit on the progenitor mass for forming a neutron star in a
binary depends on the evolutionary stage at which the star loses its
hydrogen envelope due to the interaction with its companion. This
depends on the orbital separation and the initial mass ratio.  The
lower mass limit for the formation of a neutron star in a binary may
thus be higher than for single stars.}.  Coincidence of the lower
limits of the initial primary mass for our simulation with the minimum
mass for the progenitors of neutron stars results in an underestimate
in the formation of binaries in which the secondary star experiences
a supernova.  By comparing our results with those of Portegies Zwart
\& Verbunt (1996, see their Tab.~4), who take lower mass primaries
into account, we evaluate that this affects our results with $\aplt
20$\%.

While referring the reader to PZY98 for details, we list here the most
crucial assumptions. The masses of the primaries obey a power-law with
exponent 2.5 (Salpeter $\equiv 2.35$).  The initial mass function is
normalized to the current Galactic star-formation rate of
4\,\msun\,\pyr\ (van den Hoek \& de Jong 1997).  We assume that all
stars are born in binaries with a semi-major axis up to
$a=10^6$~\rsun. The distribution was chosen to be flat in $\log
a$. Roche-lobe contact at zero age sets the minimum to the
distribution of initial separations.  The mass of the secondary is
selected to be between 0.1~\msun\ and the mass of the primary from a
flat mass ratio distribution. The eccentricity for each binary is
taken from the thermal distribution between zero and unity.

Neutron stars receive a kick upon birth.  The velocity of the kick is
taken randomly from the distribution function proposed by Hartman
(1997) and in a random direction.  Black holes (from stars initially
more massive than 40 \msun) receive a kick velocity which is scaled by
$M_{ns}/M_{bh}$.

From computations presented in PZY98 one may infer that our
conclusions concerning birth rates and \pe\ distributions for various
combinations of neutron star and companion are robust with respect to
reasonable variations in the most crucial model parameters and initial
conditions, such as the kick velocity distribution and common envelope
parameter, initial mass function and initial distribution in orbital
periods. The criteria for the stability of mass transfer and the
adopted rate of mass lost via a stellar wind and the amount of
momentum lost per unit mass, however, may affect the details of our
calculations rather significantly.

From the output of the computer simulations we select binaries with at
least one neutron star.  Within these constraints three orbital-period
ranges are considered: $P_{\rm orb} \geq 10, 100, {\rm and}\ 1000$\
days.  Table~\ref{tab:table} gives the model birth rates for the
binaries in various groups depending on the nature and the mass of the
companion of the neutron star.  Figures 1 and 2 give the probability
distributions for these binaries in the orbital period-eccentricity
plane.

\begin{figure}
\centerline{ 
\hspace{-0.98cm}
\psfig{file=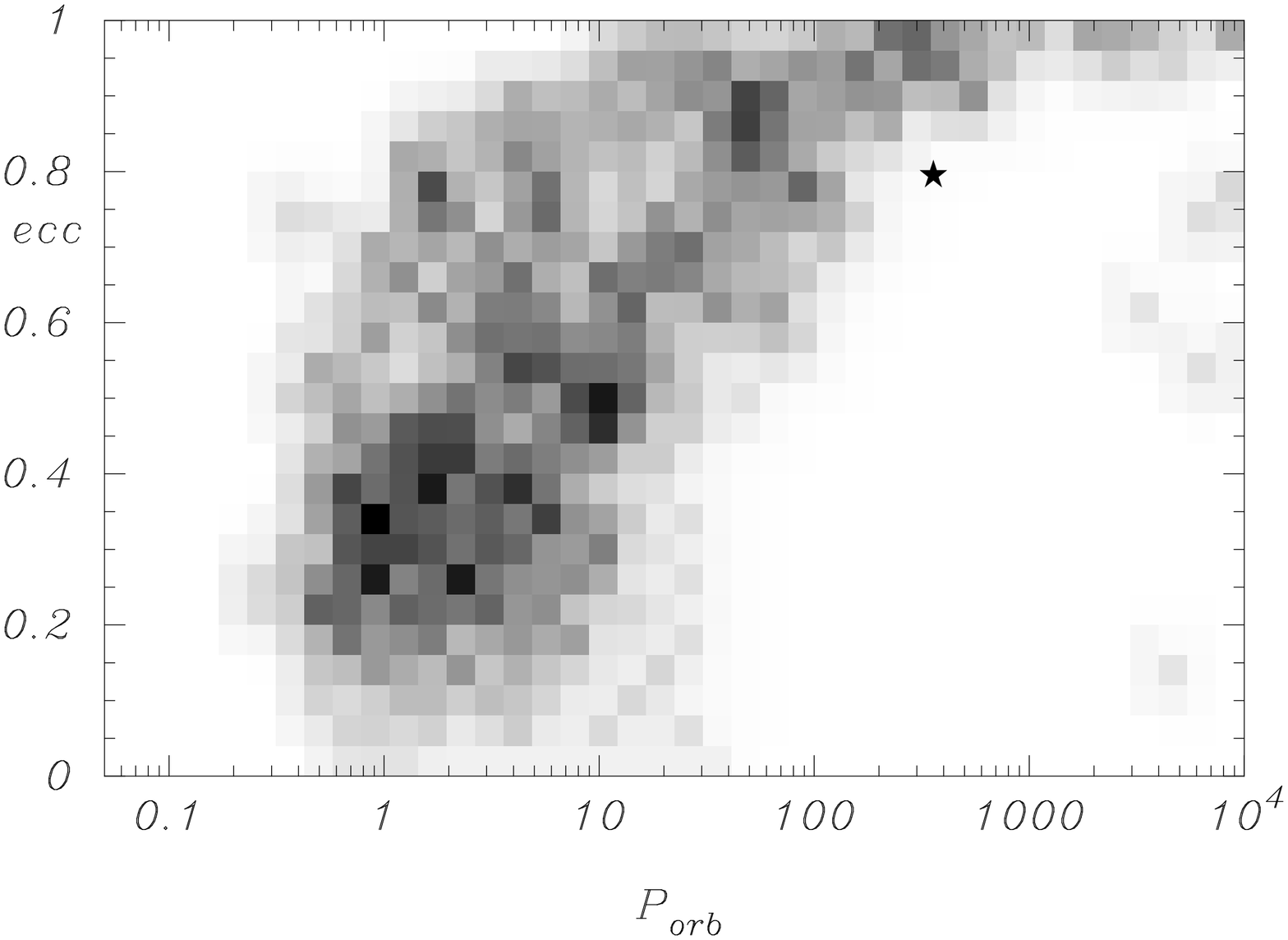,height=5cm,angle=-90}
}
\vspace{-0.61cm}
\centerline{\hspace{-0.8cm}
\psfig{file=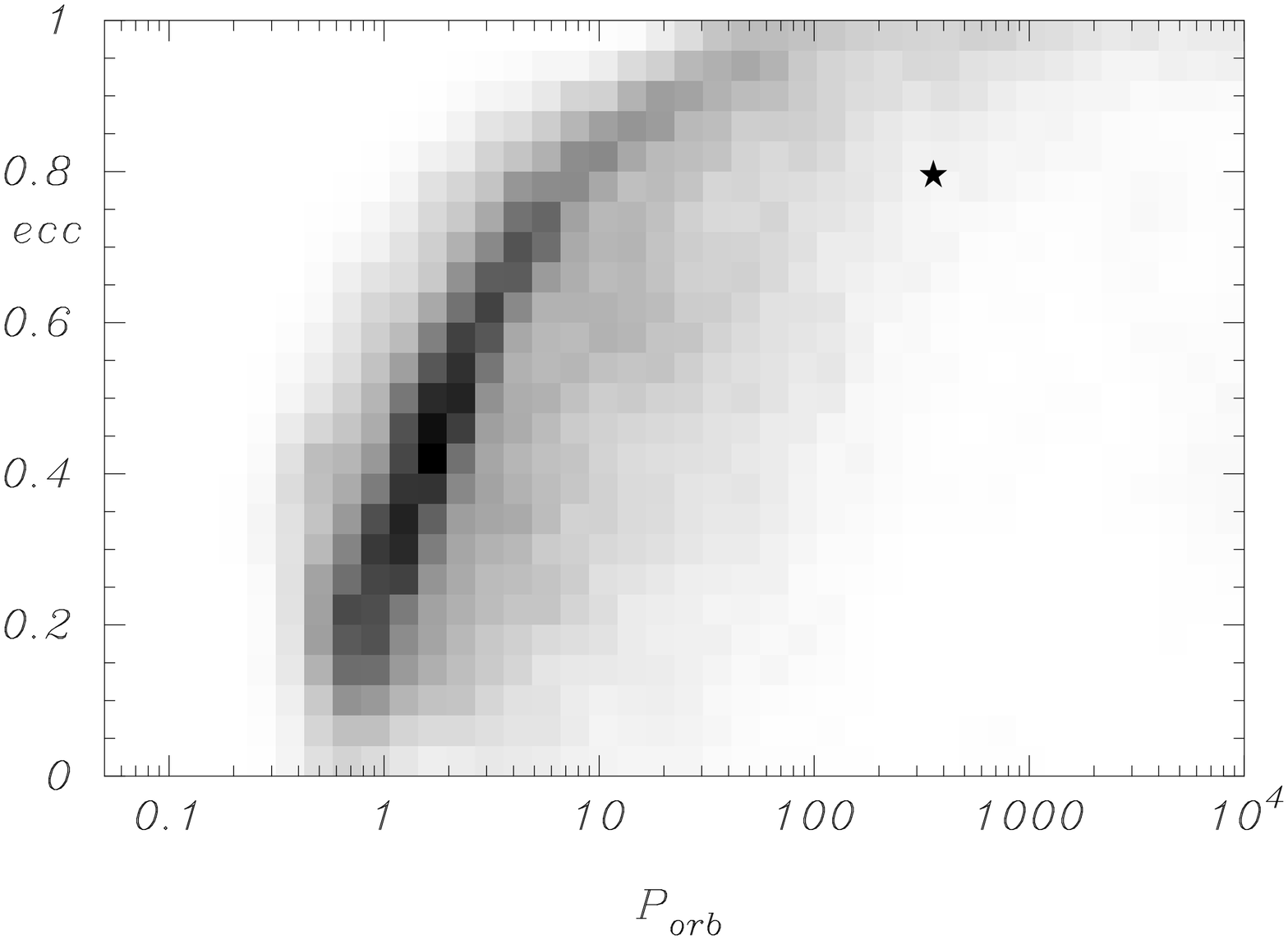,height=5cm,angle=-90}
}
\vspace{-0.61cm} 
\centerline{\hspace{-0.8cm}
\psfig{file=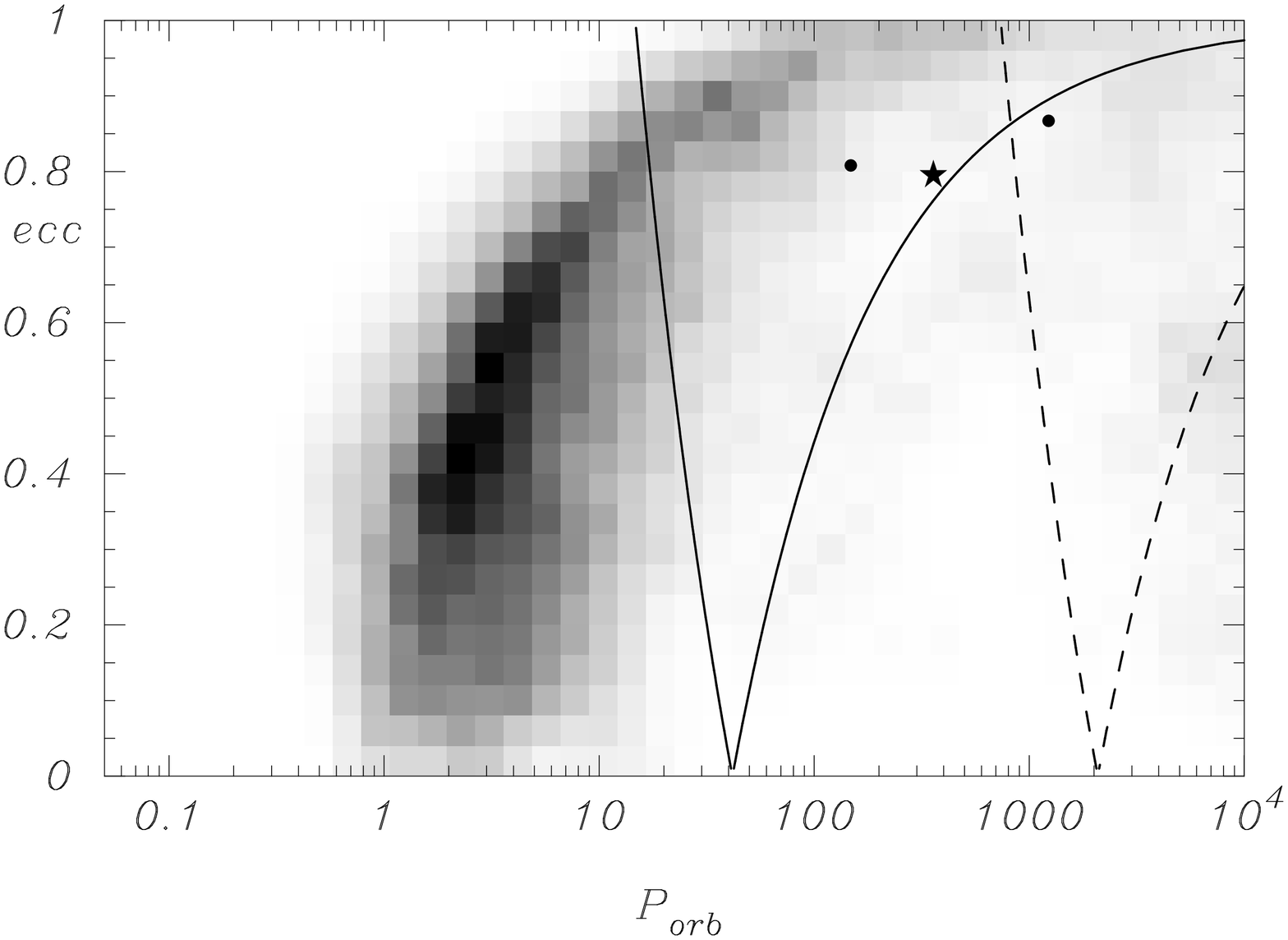,height=5cm,angle=-90}
}
\caption[]{ Probability distribution in the orbital period --
eccentricity plane for (\ns, {\it ms}) binaries with a primary mass
between 0.7\,\msun\ and 1.4\,\msun (upper panel), 1.4\,\msun\ and
5\,\msun\ (middle) and between 5\,\msun\ and 10\,\msun\ (lower panel,
see Tab.\,\ref{tab:table} for the birth rates to which the panels are
normalized).  The solid line in the lower panel is plotted using
Eq.\,(\ref{eq:ppulsar}) for a 5\,\msun\ star with a wind mass loss
rate of $10^{-9}$\msun\,yr$^{-1}$.  A binary is visible over its
entire orbit to the right of both lines. To the left of both lines,
the binary stays invisible throughout its entire orbit. In between the
left and the right part of the solid lines the binary is visible for
part of its orbit (near apocenter).  The dashed line is plotted using
a 10\,\msun\ primary with a mass loss rate of
$10^{-7}$\msun\,yr$^{-1}$.  The $\star$ symbol indicates the position
of PSR~B1820-11, and two bullets (lower panel) indicate the positions
of PSR~B1259-63 (right) and J0045-7319 (left).}
\label{fig:pe1820}
\end{figure}

The birth rates in Table~\ref{tab:table} reflect the initial
distributions of the binary parameters and the complicated
evolutionary history via several phases of mass transfer and at least
one supernova.  It is therefore not surprising that the most common
companion to a pulsar is a rather massive ($\apgt 5\msun$)\
main-sequence star in a relatively wide binary ($ 10 \aplt P_{\rm orb}
\aplt 1000$\,days, Fig. 1).  As argued above, the presence of such a
massive companion in \psr\ may be excluded by the absence of an
optical counterpart and the unlikely inclination of the orbit; hiding
a companion of $M\apgt 3\msun$\ demands $\cos i \apgt 0.95$\ (Lyne \&
McKenna 1989). The observed radio emission also makes it unlikely that
the mass of the companion to \psr\ is $\apgt 5\msun$.  Such a
companion star with a mass loss in the stellar wind of $\aplt 10^{-9}
\myr$ (which is chosen on the high side to account for the enhanced
wind mass loss for Be stars) would easily shield the radiation of the
radio pulsar along its entire orbit (see Fig.\,\ref{fig:pe1820}).

In the range of orbital periods $P_{\rm orb} \geq 100$, days the birth
rates of neutron stars accompanied either by another neutron star
(\ns), a white dwarf (\wid ), a low mass $(M \la 1.4\,\msun )$~
main-sequence star ({\em ms}) or a black hole (\bh ) are comparable
(Table \ref{tab:table}).  The relatively large population of (\bh,
\ns) systems and their distribution in orbital periods is a result of
the evolutionary history of massive stars which is dominated by
stellar wind mass loss rather than Roche lobe overflow and by the
lower kick velocities imparted to black holes.  However, the mass
function of \psr\ makes a black hole with mass exceeding several
\msun\ as a companion unlikely, and for \psrb\ such a massive
companion is excluded by the measured total mass.

\begin{table}\centering
\caption[]{ Results of the model computations for binaries with at
least one radio pulsar. The first column identifies the various binary
components.  Cols.\ (2) and (3) give the mass limits for the companion
of the pulsar (black holes have a mass $> 2\,\msun$).  Col.\ (4) gives
the total Galactic birth rate (BR) of such binaries.  Cols.\ (5), (6)
and (7) give the birth rate of binaries with an orbital period larger
than 10, 100, and 1000 days, respectively, in the units of
$10^{-5}$\,\pyr }.
\begin{flushleft}
\begin{tabular}{lrlcrrr} 
          & $m_{\rm min}$&$m_{\rm max}$ & BR
                       & $P_{\geq10}$      &      $P_{\geq100}$      &
$P_{\geq1000}$\\[5pt]
& \multicolumn{2}{c}{\unit{\msun}} & 
\multicolumn{4}{c}{\unit{$10^{-5}$\pyr}} \\ [5pt] 
(1) & (2) & (3) & (4) & (5) & (6) & (7) \\ [5pt]
(\ns, \ms)  &0.1 & 0.7& 0.5 & 0.1 & 0.0  & 0.0  \\
(\ns, \ms)  &0.7 & 1.4& 4.4 & 1.2 & 0.3  & 0.0  \\
(\ns, \ms)  &1.4 & 5.0&13.9 & 5.0 & 2.3  & 0.3  \\
(\ns, \ms)  &5.0 &10.0& 9.5 & 3.6 & 1.1  & 0.5  \\
(\ns, \ms)  &10.0& up &19.7 &17.4 &10.8  & 3.3  \\ [5pt] 
(\wid, \ns)  &1.1 & 1.4& 4.4 & 1.3 & 0.3  & 0.0  \\
(\ns, \ns)  &1.4 & 1.4& 3.4 & 0.6 & 0.2  & 0.1  \\
(\bh, \ns)  &2.0 & var.& 0.6 & 0.3 & 0.2  & 0.1  \\
\end{tabular}
\end{flushleft}
\label{tab:table}
\end{table}

\begin{figure}
\centerline{\hspace{-0.8cm}
\psfig{file=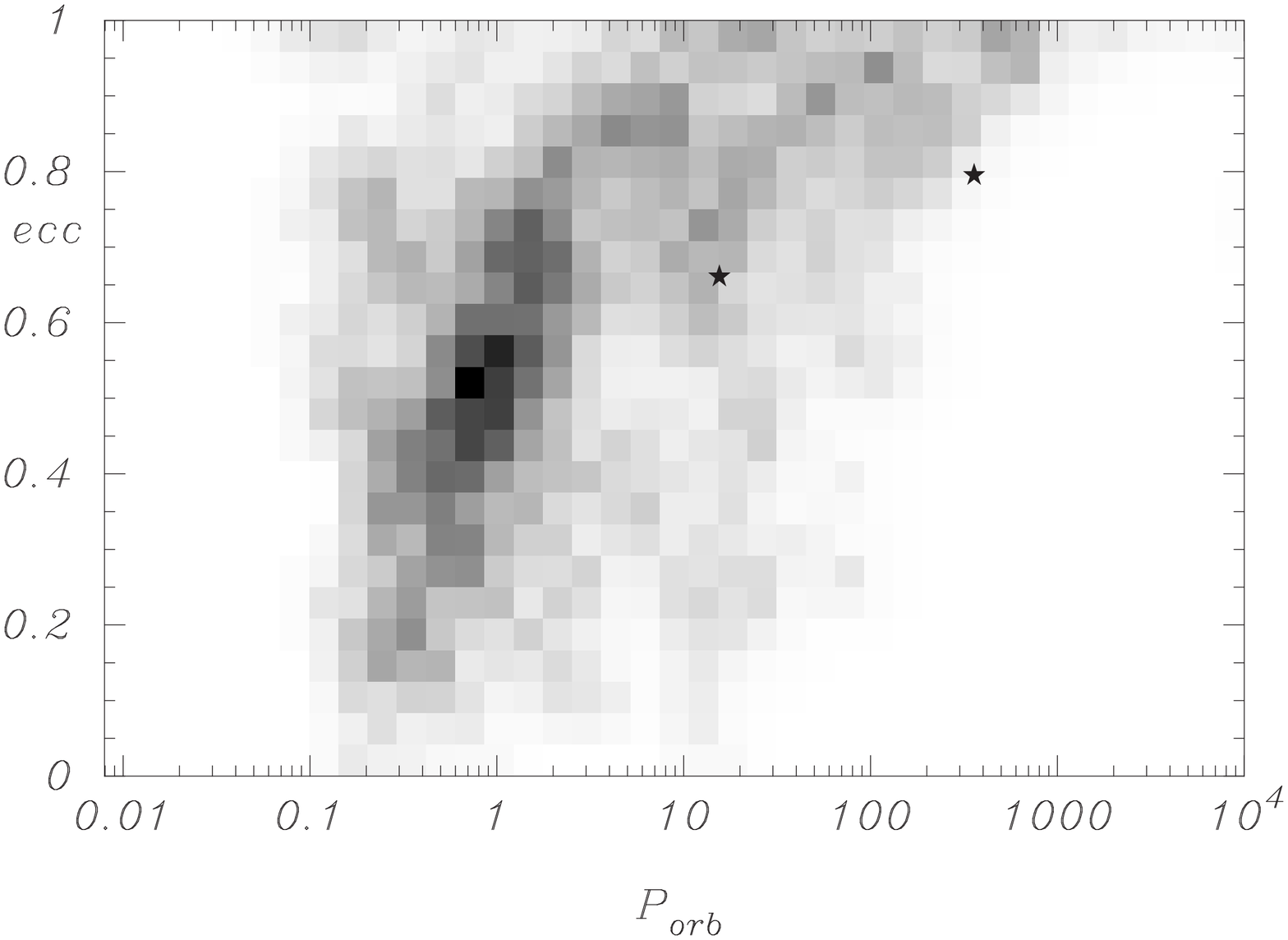,height=5cm,angle=-90}
}
\vspace{-0.61cm}
\centerline{\hspace{-0.8cm}
\psfig{file=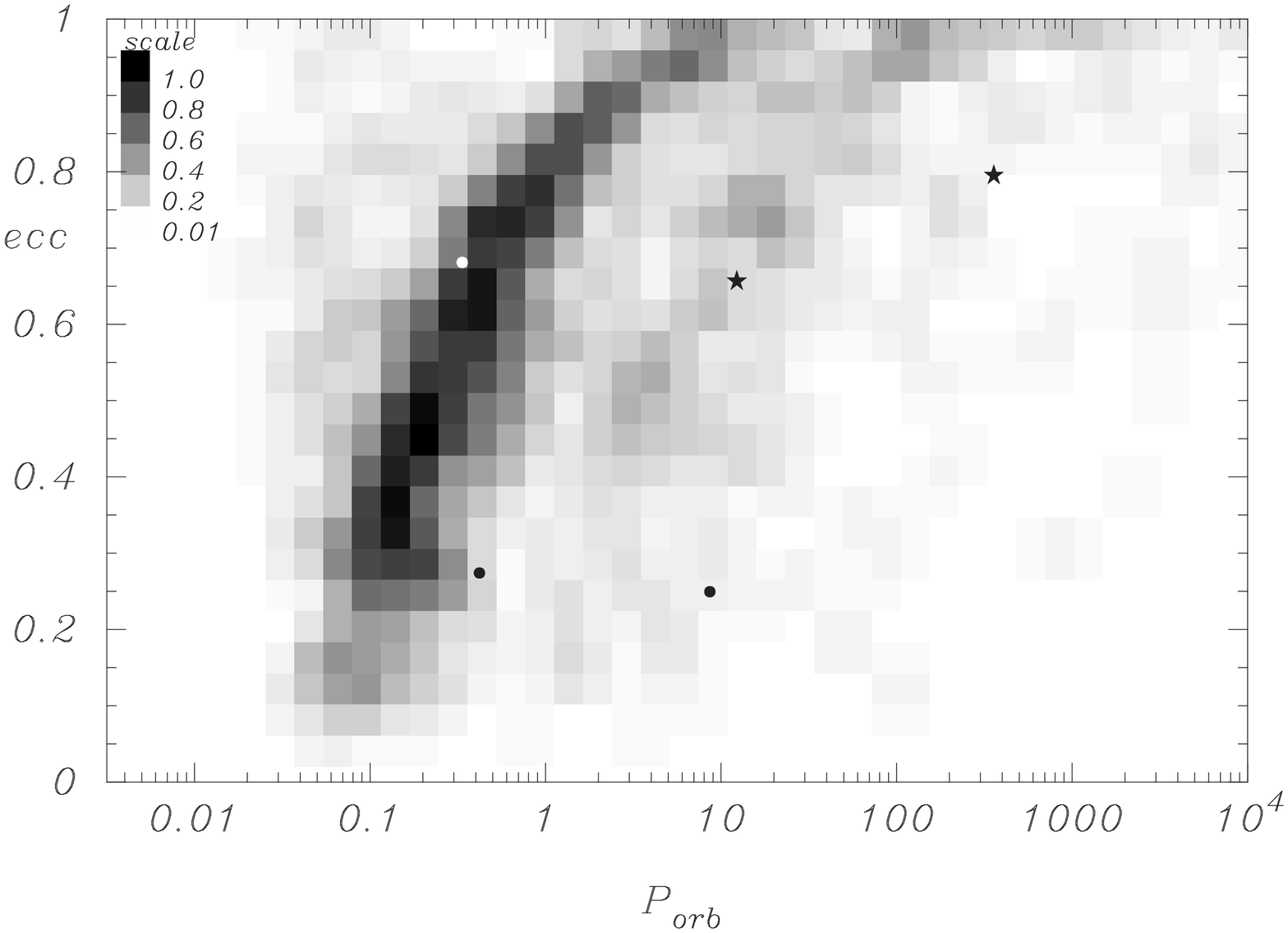,height=5cm,angle=-90}
}
\caption{Probability distributions for binaries which contain a young
neutron star and a white dwarf (top panel) or a neutron star (lower
panel) as a companion star.  
A figure for (\bh, \ns) is
not provided because of the small number of data points.  The $\star$\ symbols
identify the positions of \psr\ (right) and \psrb\ (left).  Dots in
the lower panel give the positions of the known recycled pulsars with
a neutron star companion. 
}
\label{figallpe}
\end{figure}

Figure~\ref{figallpe} compares the probability distributions for
binaries with a radio pulsar which is accompanied by a white dwarf
(upper panel) and an older neutron star (lower panel) in the orbital
period - eccentricity diagram.

Our model calculations reveal that the binaries in which a young
pulsar is accompanied by a massive white dwarf are generally rather
eccentric ($e \sim 0.6$) and that their orbital periods are
distributed with more or less equal probability in $\log P_{\rm orb}$
between a few hours and about one hundred days.  In Fig.\,
\ref{fig:pe1820} \psr\ appears at the edge of the area with highest
probability, while the orbital parameters of \psrb\ fall nicely in the
densely populated area of the \pe\ diagram. This suggests, together
with the rather high birth rate of such binaries, that \psrb\ may well
contain a white dwarf as a companion star.  Recently van Kerkwijk \&
Kulkarni (1999) found evidence that \psrb\ is accompanied by a white
dwarf with a mass of about 1.2\,\msun.

\psr\, fits rather ill in either of the \pe\ diagrams for (\ns, \ns),
(\wid, \ns), and low mass or moderate mass (\ns, \ms) binaries. 
The relative birthrates for a
population of binaries with long orbital periods increases 
by increasing the efficiency of the deposition of orbital energy into
common envelopes during the spiral-in phase (see PZY98). 

However, this would require us to relax the decisive criterion why we
chose this specific set of model parameters; the coincidence of having
the other high mass binary pulsars (PSR J1518+49, B1534+12, and
B1913+16) in the most densely populated areas in the \pe\ diagram for
(\ns, \ns) binaries.  There is no evidence for a strong selection
against the discovery of a young pulsar which is accompanied by a dead
pulsar or by a massive white dwarf with an orbital period $\apgt
10$\,days (Johnston \& Kulkarni 1991). The higher birth rate of
binaries in which the radio pulsar is accompanied by a low mass
main-sequence star and the better coincidence in the \pe\ diagram
makes this possibility very attractive.

If the companion of \psr\ is indeed a low-- or moderate mass
main-sequence star it may be a precursor of a system similar to GX~1+4
= V2116 Oph, which shows the X-ray features of an accreting neutron
star and which is identified with a symbiotic star (Davidsen, Malina,
\& Bowyer 1977). The orbital period of GX~1+4 is $\apgt 100$\, days
(Chakrabarty \& Roche 1997). The low birth rate of such binaries
(Table\, \ref{tab:table}) together with the short life time of a red
giant explains the paucity of such X-ray binaries.

\section{Conclusion}

We have studied the nature of the companions of young radio pulsars in
eccentric orbits.  Our calculations reveal that the possibility that a
young pulsar is accompanied by a white dwarf of mass $\apgt 1.1\msun$
cannot be neglected.  The birth rate of such binaries is higher than
the birth rate of of binaries in which a young pulsar is accompanied
by a dead pulsar (i.e.: an old neutron star).

We argue therefore that \psr\ may be accompanied by a massive white
dwarf, however the probability that its companion is a main-sequence
star with a mass $\aplt 5\msun$ (or even another neutron star) is
similar to the probability of being a white dwarf.

Our calculations confirm the observation of van Kerkwijk \& Kulkarni
(1999) in the sense that it is likely that \psrb\ is accompanied by a
white dwarf with a mass $\apgt 1.2\,\msun$.

\section*{Acknowledgments}
We thank Thereasa Brainerd, Ramach Ramachandran and the anonymous
referee for reading the manuscript.  We acknowledge the kind
hospitality of the Astronomical Institute 'Anton Pannekoek', Meudon
Observatory and the Univeristy of Tokyo where a part of this work was
done.  This work is supported by NWO Spinoza grant 08-0 to
E.~P.~J.~van den Heuvel, RFBR Grant 96-02-16351, the JSPS fellowship
to SPZ and by NASA through grnat number HF-01112.01-98A from the Space
Telescope Science Institute, which is operated by the Association of
Universities for Research in Astronomy, Inc., under NASA contract
NAS\, 5-26555.

\label{lastpage}

\begin{thebibliography}{}
\bibitem{}
Bell J. F., Bessell M. S., Stappers B. W., Bailes M., Kaspi,
V. M., 1995, ApJ, 447, L117

\bibitem{} Blaauw A., 1961, BAIN, 15, 265




\bibitem{} Chakrabarty D., Roche P., 1997, ApJ, 489, 254


\bibitem{}
Grove J. E., Tavani M., Purcell W. R., Johnson W. N., Kurfess J. D.,
Strickman M. S., Arons J., 1995, ApJ, 447, L113

\bibitem{}
Davidsen A., Malina R., Bowyer S., 1977, ApJ, 211, 866


\bibitem{}
Hartman J. W., 1997, A\&A, 322, 127


\bibitem{}
Illarionov A. F., Sunyaev R. A., 1975, A\&A, 39, 185

\bibitem{} Johnston H. M., Kulkarni S. R., 1991, ApJ, 368, 504 

\bibitem{}
Johnston, S., Manchester R. N., Lyne A. G., 
Bailes M., Kaspi V. M., Qiao G.,  D'Amico N.,
1992, ApJ, 387, L37

\bibitem{}
Johnston S., Manchester R. N., Lyne A. G., 
Nicastro L., Spyromillo J., 1994, MNRAS, 268, 430



\bibitem{}
Kaspi V. M., Tauris T. M., Manchester R. N., 1996, ApJ, 459, 717.

\bibitem{} 
Koester D., Reimers, D., 1996, \aa, 313, 810


\bibitem{} Kulkarni S. R., 1988, Adv. Space Research 8, 343


\bibitem{}
Lyne A. G., McKenna J. M., 1989, Nat, 340, 367

\bibitem{}
McConnell D., McCulloch P. M., Hamilton P. A., 
Ables J. G., Hall
P. J., Jacka C. E., Hunt A. J. 
\etal, 1991, MNRAS, 249, 654 

\bibitem{}
Melatos A, Johnson S., Melrose D.B., 1995, MNRAS, 275, 381


\bibitem{}
Monet, D., Bird, A., Canzian, B. \etal, A Catalog of Astrometric Standards,
USNO-A, v1.0, U.S. Naval Observatory   

\bibitem{}
Phinney E. S., Verbunt F., 1991, MNRAS, 248, 21




\bibitem{}
Portegies Zwart S. F., Verbunt F.,  1996, A\&A, 309, 179

\bibitem{}
Portegies Zwart S. F., Yungelson L. R., 1998, A\&A, 332, 173,







\bibitem{}
Ritossa, C., Garcia-Berro, E., Iben, I. Jr., 1996, \apj, 460, 489 

\bibitem{}
Stokes G., Taylor J., Dewey R., 1985, ApJ, 294, L91




\bibitem{}
Taylor J. H., Manchester R. N., Lyne A. G., 1993, ApJSS, 88, 529

\bibitem{}
Thorsett S. E., Chakrabarty D., 1999, ApJ 512, 288

\bibitem{}
Tutukov A. V., Yungelson L. R., 1993, SvA, 37, 411


\bibitem{}
van den Hoek B., de Jong T., 1997, A\&A, 318, 231

\bibitem{}
van Kerkwijk, M. H., Kulkarni, S. R., 1999, astro-ph/9901149




\end{thebibliography}
\end{document}